\documentclass[aps,twocolumn,showpacs,showkeys,superscriptaddress,amsmath,amssymb,nofootinbib,floatfix,prc]{revtex4-1}

\usepackage{graphicx}
\usepackage{subfigure}
\usepackage{bm}

\makeatletter

\begin{document}

\title{Charge conservation and the shape of the ridge of two-particle correlations in  relativistic heavy-ion collisions}

\author{Piotr Bo\.zek}
\affiliation{The H. Niewodnicza\'nski Institute of Nuclear Physics, Polish Academy of Sciences, PL-31342 Krak\'ow, Poland}
\affiliation{Institute of Physics, Rzesz\'ow University, PL-35959 Rzesz\'ow, Poland}

\author{Wojciech Broniowski}
\affiliation{The H. Niewodnicza\'nski Institute of Nuclear Physics, Polish Academy of Sciences, PL-31342 Krak\'ow, Poland} 
\affiliation{Institute of Physics, Jan Kochanowski University, PL-25406~Kielce, Poland}

\date{16 April 2012}

\begin{abstract}
We demonstrate that in the framework of the event-by-event hydrodynamics followed by statistical hadronization, 
the proper charge conservation in the mechanism of hadron production provides the crucial non-flow 
component and leads to agreement 
with the two-dimensional two-particle correlation data in relative azimuthal angle and pseudorapidity at soft transverse momenta ($p_T<2$~GeV). 
The fall-off of the same-side ridge in relative pseudorapidity follows from the fact that a pair of particles with 
balanced charges is emitted from the same fluid element, whose collective velocity
collimates the momenta of the pair. We reproduce basic experimental features of the two-dimensional correlation function, such as the dependence on 
the relative charge and centrality, as well as the related charge balance functions and the harmonic flow coefficients as functions of the relative 
pseudorapidity.
\end{abstract}



\maketitle

Two-particle correlation functions in the relative angle $\Delta \phi$ and pseudorapidity $\Delta \eta$ are valuable tools to study collective flow and the mechanism of 
particle emission in relativistic heavy-ion collisions. The harmonic components of the collective flow are visible in the dihadron
correlation function as two {\em ridge} structures on the same ($\Delta \phi \simeq 0$) and away ($\Delta \phi \simeq \pi$) 
sides \cite{Takahashi:2009na,*Luzum:2010sp}. There is, however, an on-going 
discussion concerning the puzzling nature of the same-side ridge~\cite{Agakishiev:2011pe}. 
While it is commonly accepted that the collective harmonic flow \cite{Takahashi:2009na,Luzum:2010sp} determines the profile in $\Delta \phi$ 
for large pseudorapidity separations, 
up to now the shape of these structures in $\Delta \eta$, in particular the 
rather fast fall-off of the same-side ridge, remains an object of active debate, with arguments that the presence 
of (mini)jets \cite{Trainor:2011cq} is essential to explain the phenomenon and that the 
applicability of hydrodynamics, reproducing numerous other features of the heavy-ion data, is at stake.
Thus the issue is of great importance for the fundamental understanding of relativistic heavy-ion collisions. 
Other attempts to explain the nature of the ridge can be found in Refs.~\cite{Gavin:2008ev,*Gavin:2011gr}.

In this Letter we show that two basic features of the two-particle correlations get a quantitative explanation via the {\em charge balance} mechanism of particle emission:
1)~the shape of the same-side ridge in $\Delta \eta$, and 2)~the  difference  between  the correlation functions for like- and unlike-sign particles. Thus we explain the ridge puzzle in a natural way, 
amending the (event-by-event, 3+1-dimensional, viscous) hydrodynamics with the local {\em charge-conservation} 
mechanism in the statistical 
hadronization occurring after the hydrodynamic evolution. 
This important {\em charge balancing} \cite{Bass:2000az,Jeon:2001ue,Bozek:2003qi,Aggarwal:2010ya}, 
simply stating that the hadron production conserves locally the charge, 
is an otherwise well-known and measured feature.


\begin{figure}[b]
\begin{center}
\vspace{-12mm}
\includegraphics[width=0.5 \textwidth]{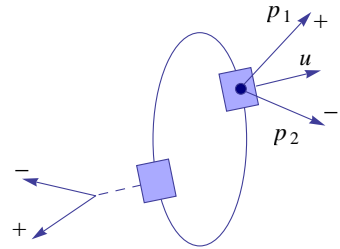} 
\end{center}
\vspace{-18mm}
\caption{(Color online) A schematic view of the charge balancing mechanism, producing pairs of particles 
with opposite charges. The rectangles indicate
fluid elements moving outward with a collective velocity $u$. The dot indicates the space-time location of the emission of the pair of 
opposite-charge particles of momenta $p_1$ and $p_2$. The dashed line represents a neutral resonance, decaying into a pair particles.
\label{fig:cartoon}} 
\end{figure}

\begin{figure*}[tb]
\begin{center}
\includegraphics[width=0.4 \textwidth]{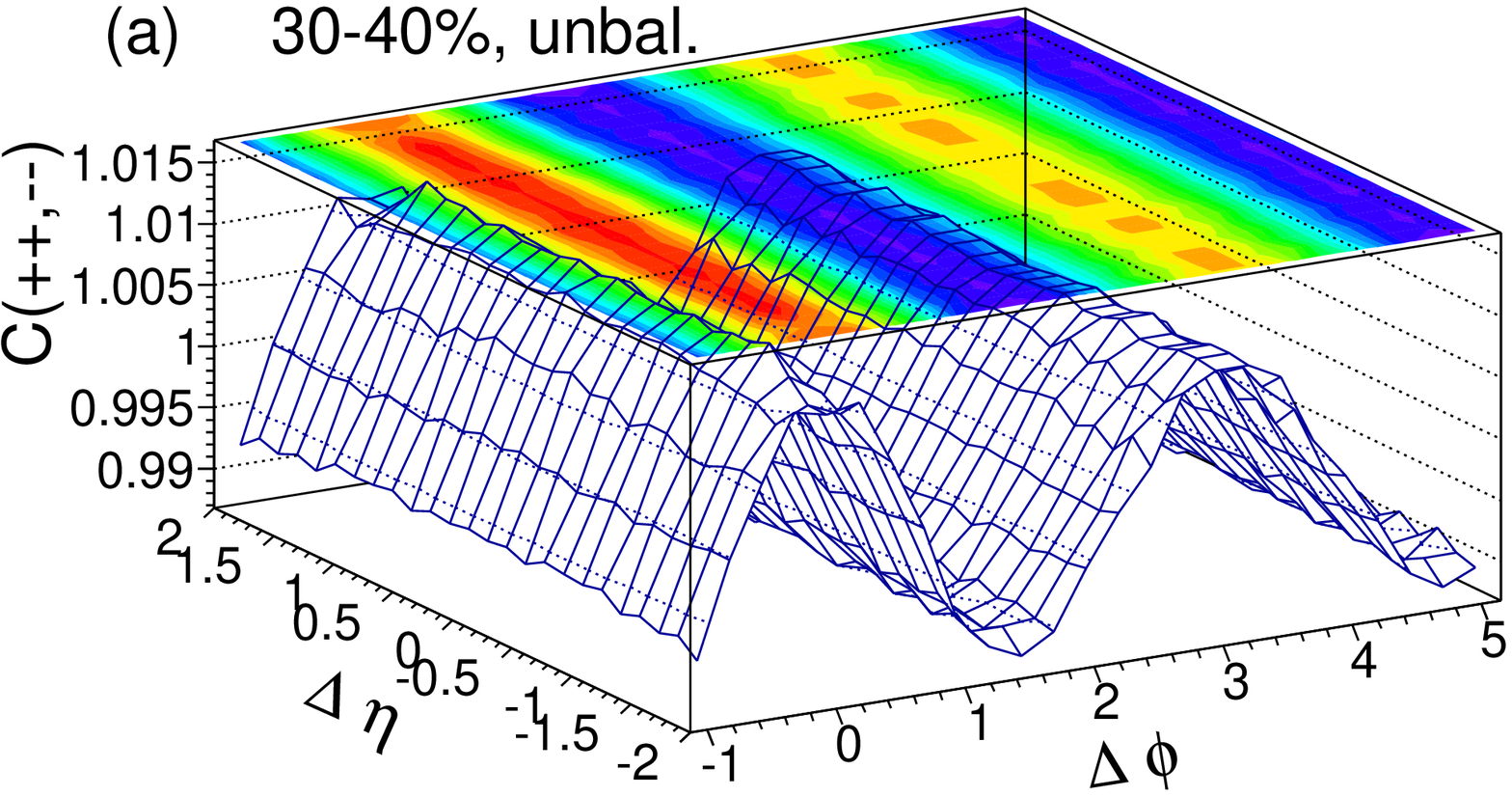} 
\includegraphics[width=0.4 \textwidth]{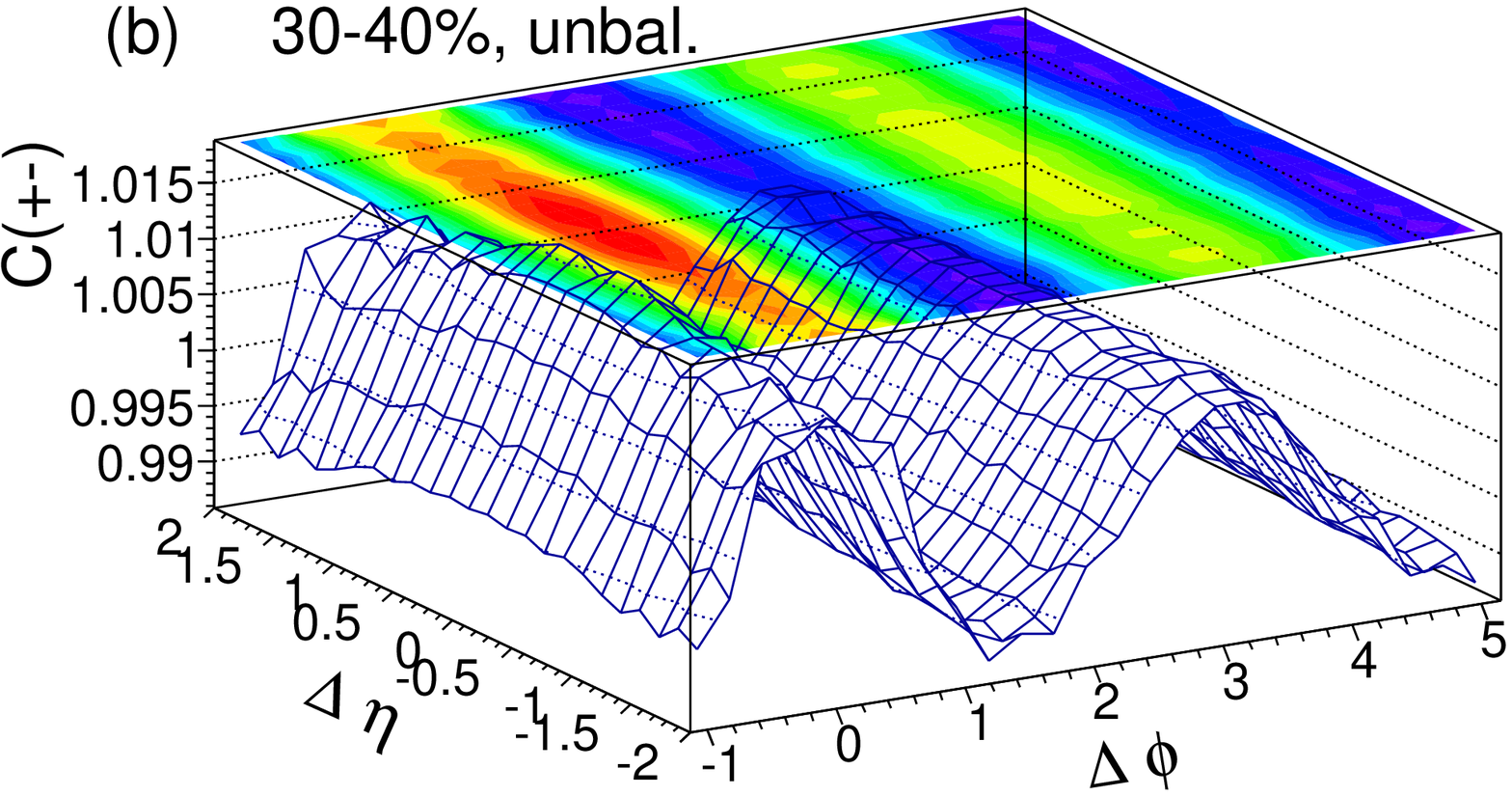} \\ \vspace{-8mm}
\includegraphics[width=0.4 \textwidth]{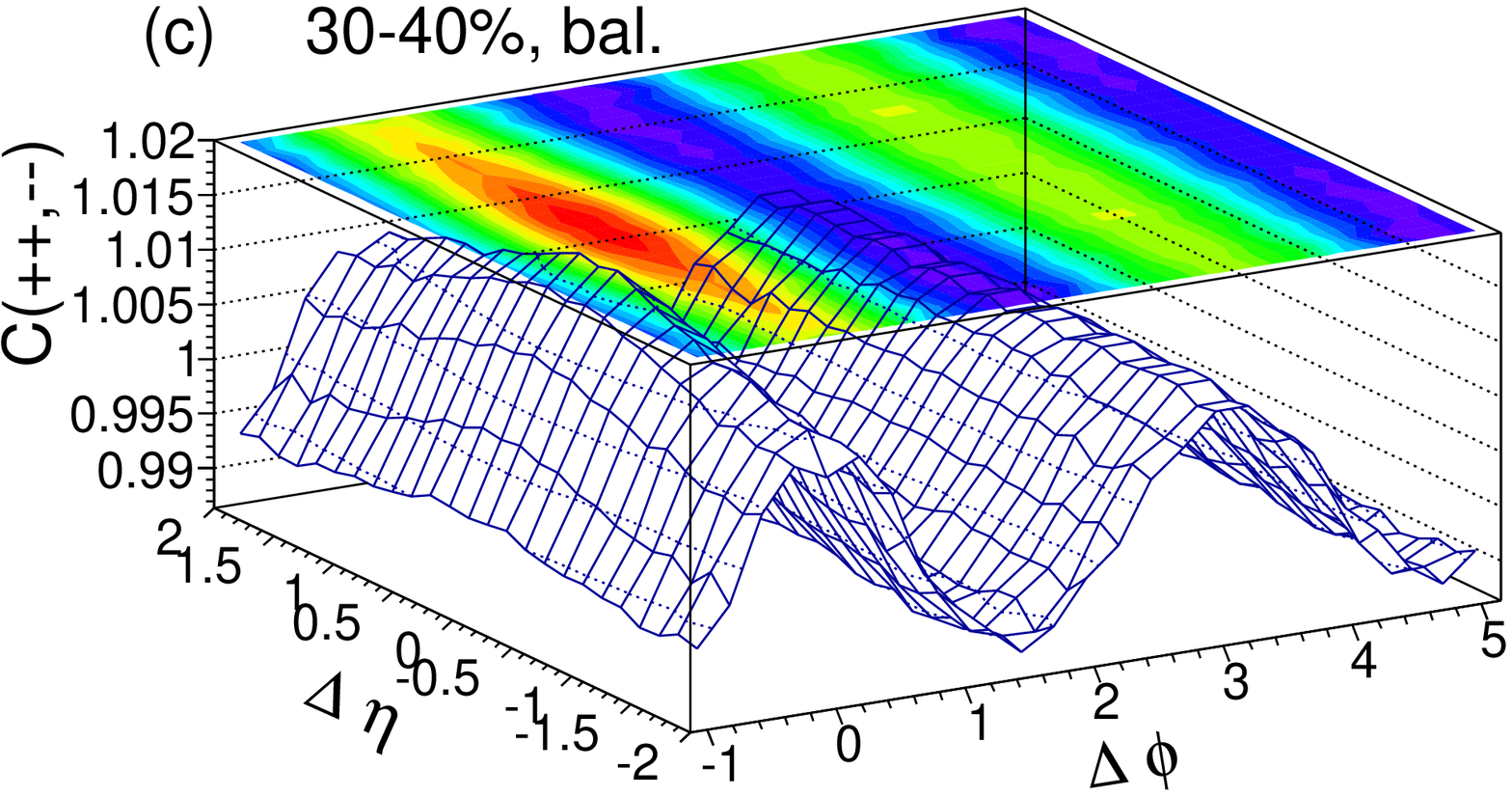} 
\includegraphics[width=0.4 \textwidth]{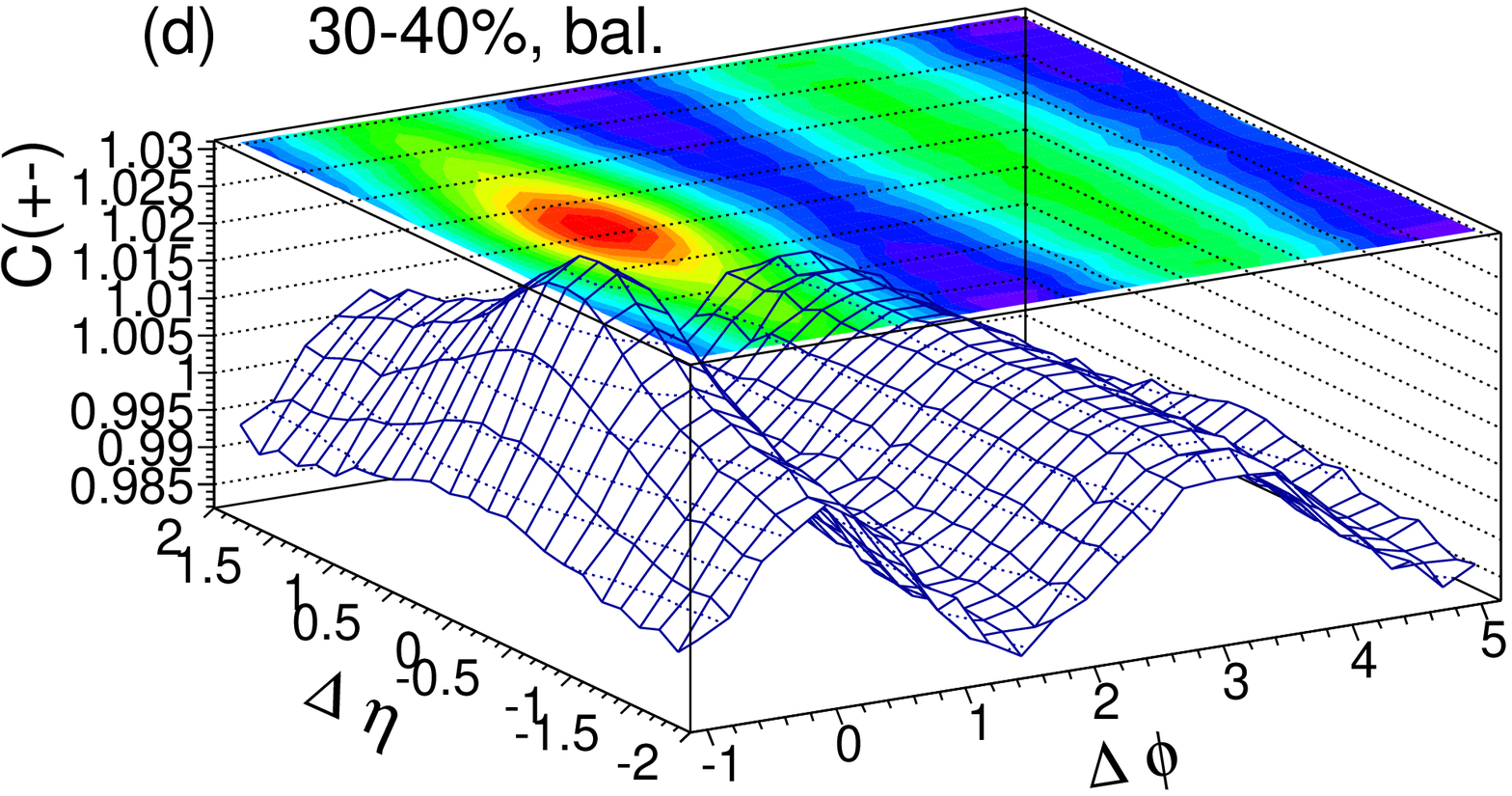} \\
\end{center}
\vspace{-8mm}
\caption{(Color online) The correlation function $C$ for like-sign (a,c) and opposite-sign (b,d) pairs. 
Panels (a,b) and (c,d) correspond to absent and present direct charge balancing, respectively. Inclusion of charge balancing 
sharpens the peak around $\Delta \eta = \Delta \phi =0$ and causes the desired 
fall-off of the same-side ridge (centrality $30-40$\%, 
$T_f=140$~MeV, $0.2 < p_T < 2$~GeV). 
\label{fig:C02}} 
\end{figure*}  

The results presented in this work concern ``soft physics'' (typically with 
the transverse momentum of all particles $p_T < 2$~GeV) and {\em unbiased} correlations, where
the kinematic cut on both particles is the same. 
The relevant correlation function is determined as
\begin{eqnarray}
C(\Delta \eta, \Delta \phi) = {N^{\rm pair}_{\rm real}(\Delta \eta, \Delta \phi)}/{N^{\rm pair}_{\rm mixed}(\Delta \eta, \Delta \phi)}, \label{eq:def}
\end{eqnarray}
where $N^{\rm pair}_{\rm real, mixed}(\Delta \eta, \Delta \phi)$ denote the two-dimensional distributions of pairs of particles with relative pseudorapidity 
$\Delta \eta$ and azimuth $\Delta \phi$, obtained from the real and mixed events, respectively.
Our 
approach consists of using {\tt GLISSANDO} \cite{Broniowski:2007nz} to generate the Glauber-model 
initial condition, then running event-by-event 3+1D hydrodynamics with shear and bulk viscosities 
\cite{Bozek:2011if,*Bozek:2012fw}, and finally 
carrying out the statistical hadronization with {\tt THERMINATOR} \cite{Chojnacki:2011hb} 
at the freeze-out temperature $T_f$. 
Our simulations incorporate the kinematic cuts of the 
STAR experiment, with $|\eta| < 1$, appropriate $p_T$ cuts specified later, as well as the detector 
efficiency at the level of 
90\%, estimated to hold for the registered charged particles in STAR. For simplicity, we set all chemical potentials 
at freeze-out to zero, which is a good approximation at RHIC. Other, more technical details of our approach may be
found in Ref.~\cite{Bozek:2012fw}).

The observed charge balance functions can be explained assuming that  opposite charge pairs are created towards
the end of the evolution \cite{Jeon:2001ue,Bozek:2003qi}.
To implement this mechanism 
in a simple model way but with a realistic hydrodynamic flow 
(we call it direct charge balancing), 
we enforce that the same-species charged hadron-antihadron pairs are produced at the same space-time 
location $x$ (see Fig.~\ref{fig:cartoon}). The hadron momenta $p_1$ and $p_2$ are determined independently according 
to the Cooper-Frye formula. The fact that the fluid element moves with a collective
velocity $u^\mu(x)$ causes a certain-degree of collimation of the momenta of the produced pair. 
An additional balancing mechanism comes from the 
decays of neutral resonances (see Fig.~\ref{fig:cartoon}). The correlations induced by balancing are of a 
non-flow character, i.e., cannot be obtained 
by the folding of single-particle distributions containing the collective flow.

\begin{figure*}[tb]
\begin{center}
\includegraphics[width=0.35 \textwidth]{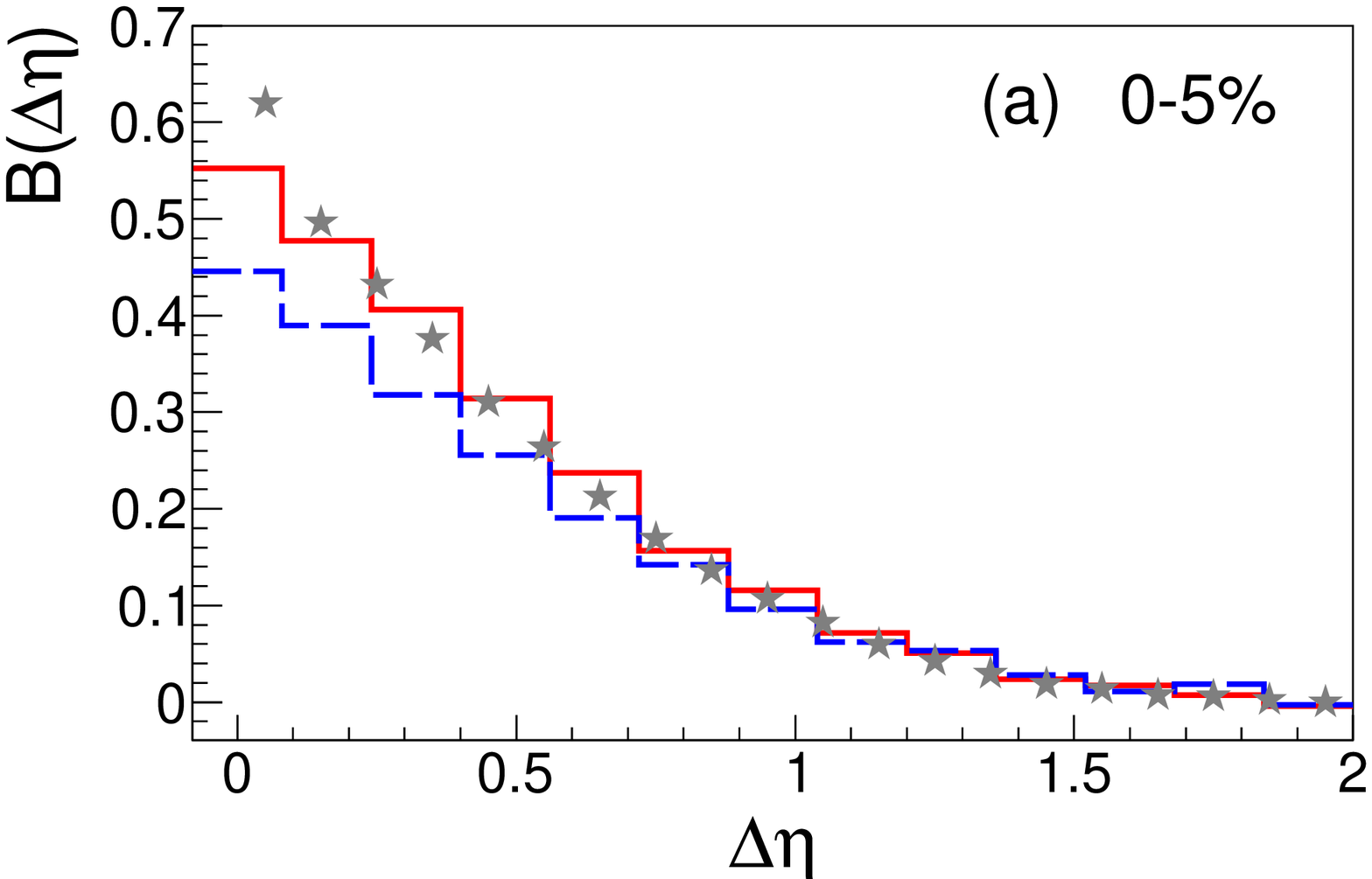} \hspace{-7mm}
\includegraphics[width=0.35 \textwidth]{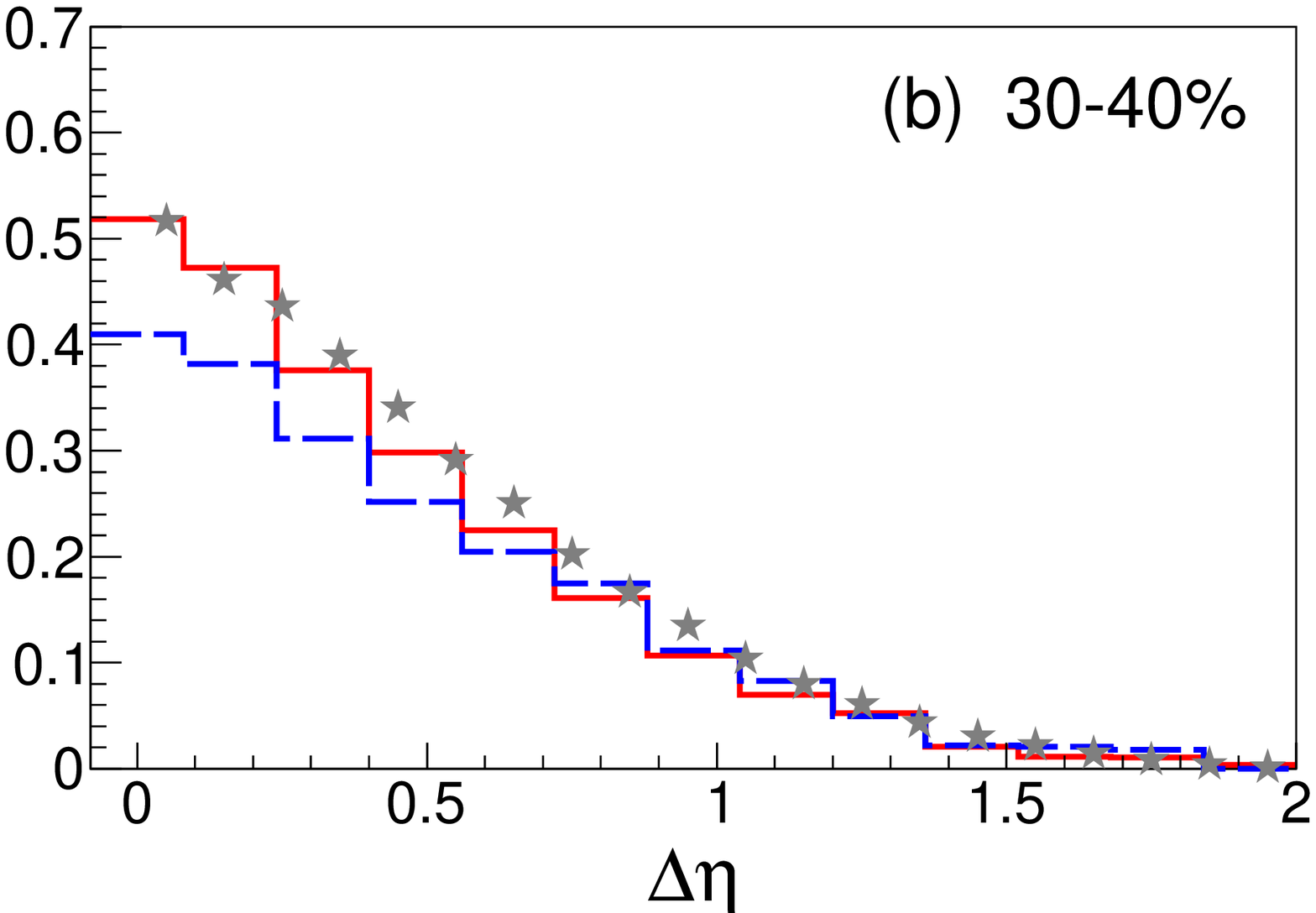} \hspace{-7mm}
\includegraphics[width=0.35 \textwidth]{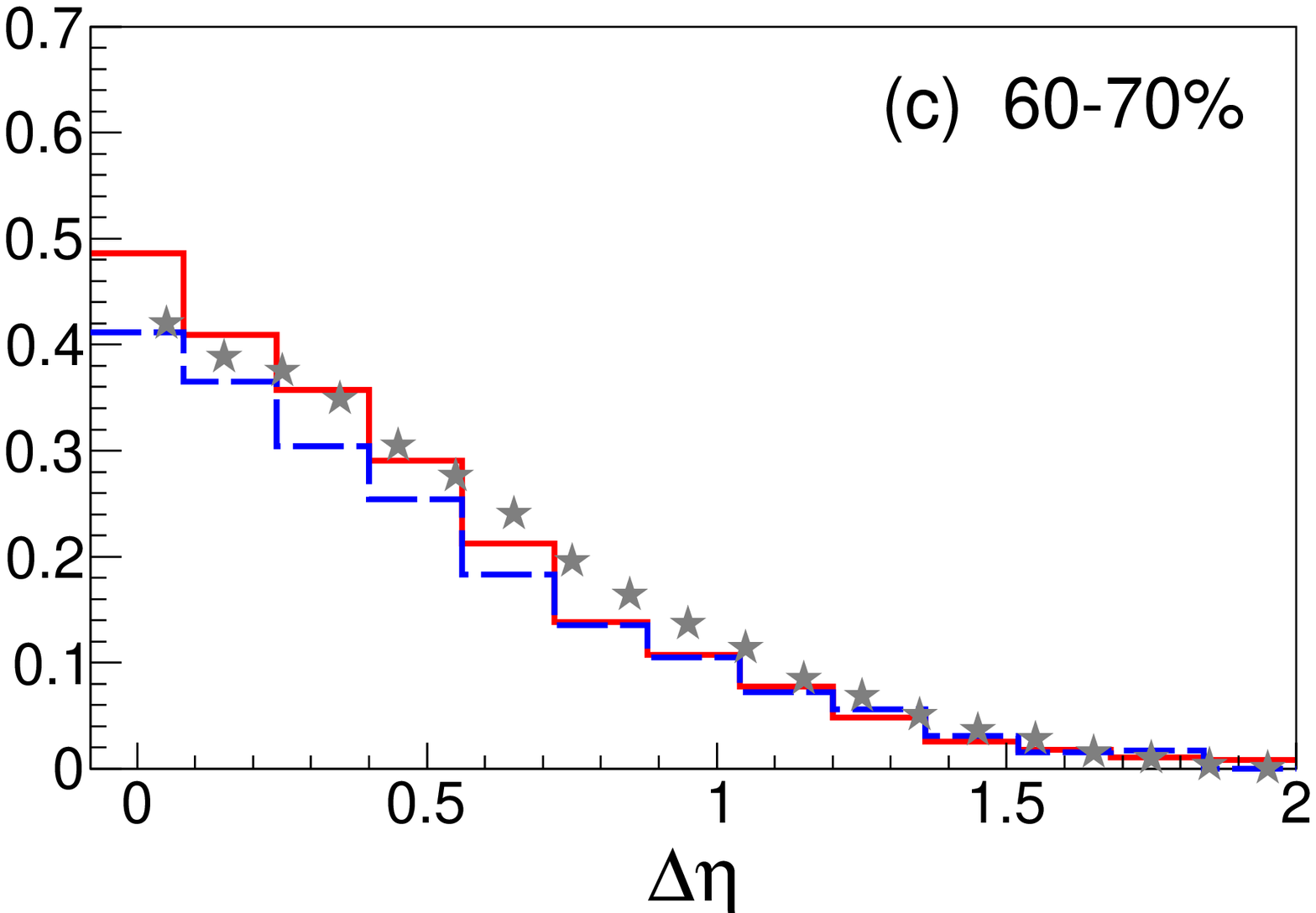} \\
\end{center}
\vspace{-7mm}
\caption{(Color online) The charge balance function for $T_f=140$~MeV (solid lines) and $150$~MeV (dashed lines). The stars indicate the 
STAR measurement at $\sqrt{s_{NN}}=200$~GeV \cite{Aggarwal:2010ya} ($0.2<p_T<2$~GeV, efficiency 90\%).
\label{fig:bal}} 
\end{figure*}

\begin{figure*}[tb]
\begin{center}
\vspace{-5mm}
\includegraphics[width=0.357 \textwidth]{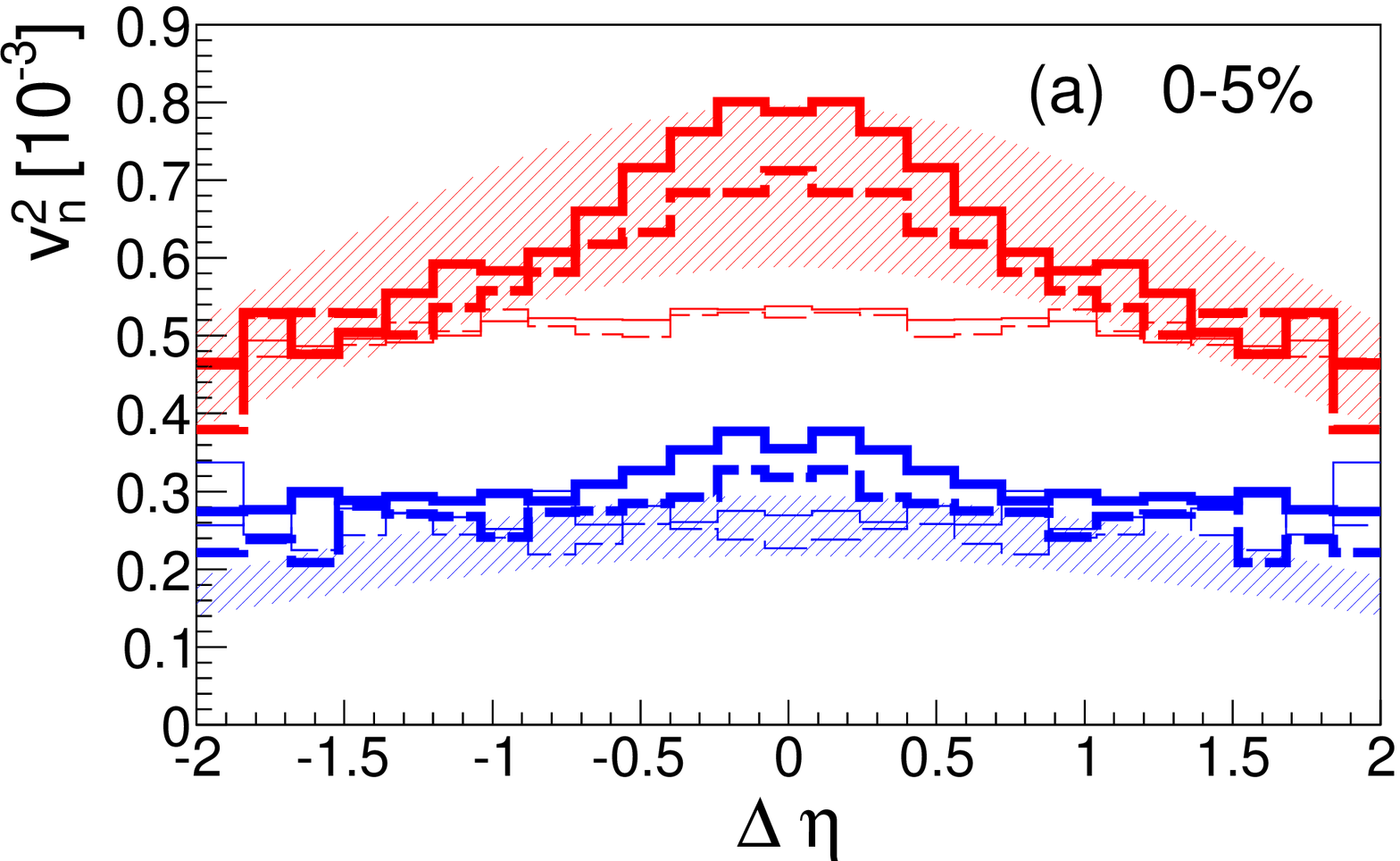} \hspace{-9mm}
\includegraphics[width=0.357 \textwidth]{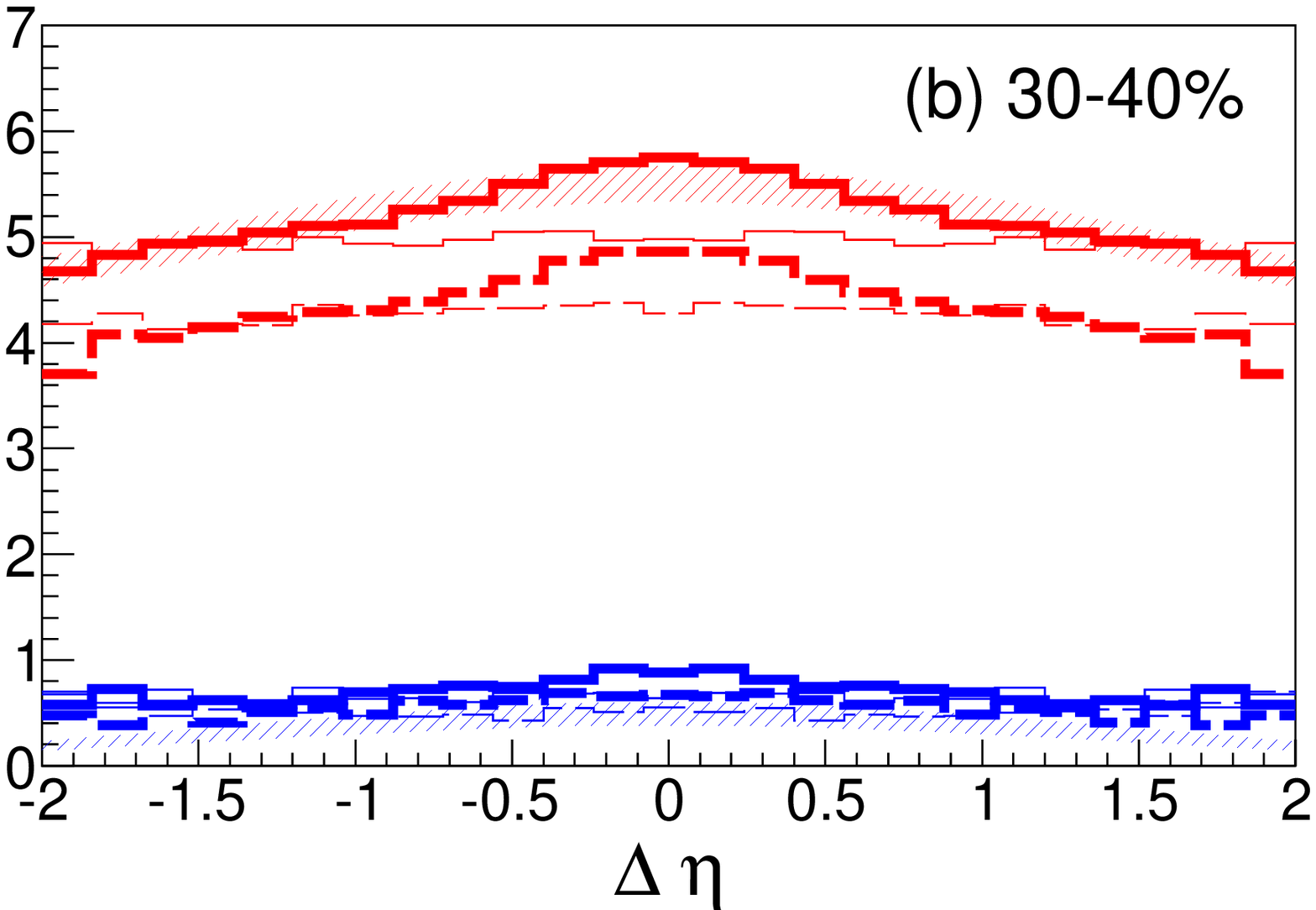} \hspace{-9mm}
\includegraphics[width=0.357 \textwidth]{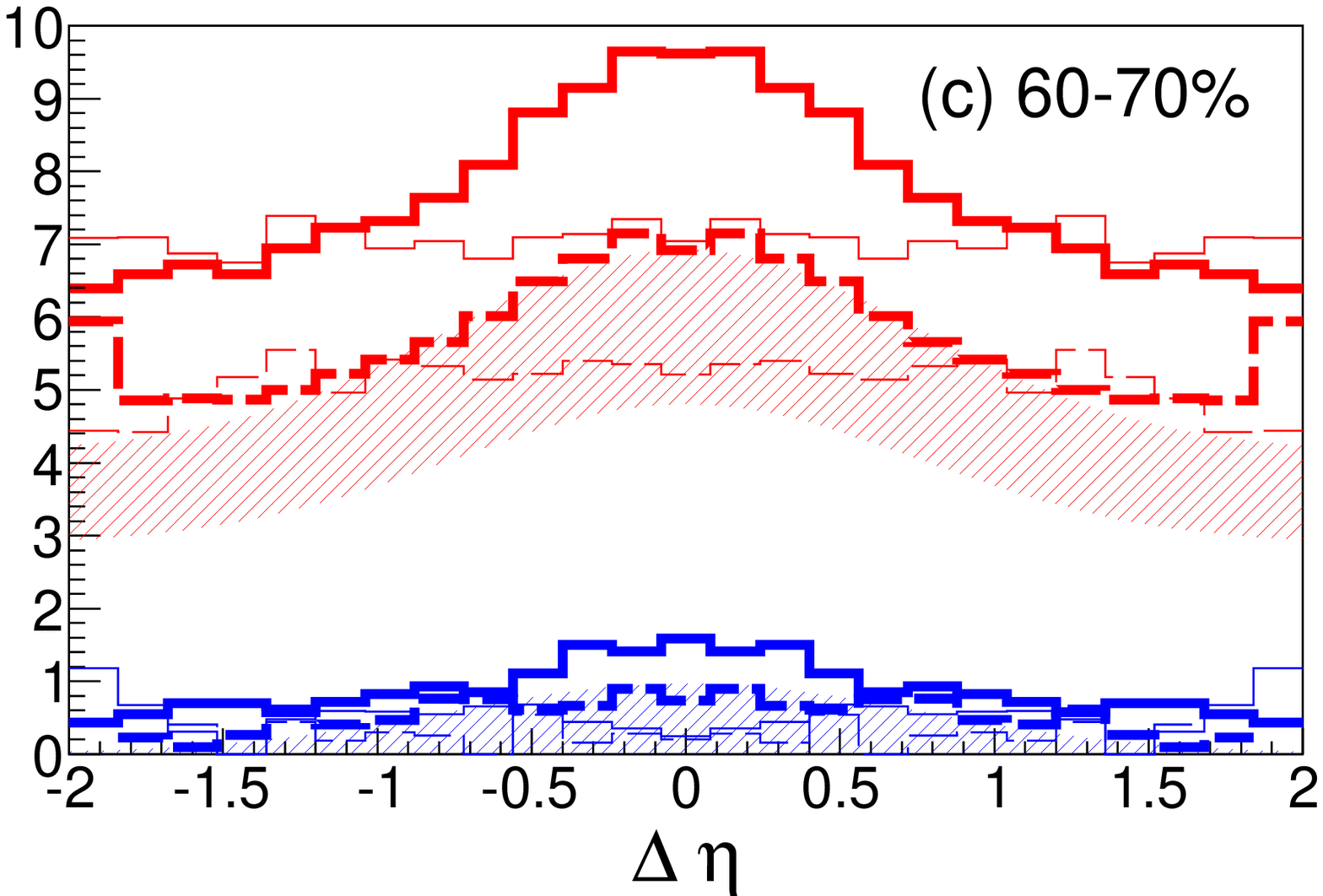} \\
\end{center}
\vspace{-7mm}
\caption{(Color online) The flow coefficients $v_2^2$ (top lines) and $v_3^2$ (bottom lines). 
Simulations with direct charge balancing are drawn with
thick solid ($T_f=140$~MeV) and dashed lines ($T_f=150$~MeV), while the 
corresponding reference simulations without 
direct charge balancing are drawn with thin lines.  The dashed bands 
are extracted from the fits to experimental data reported by STAR in Table~I of Ref.~\cite{Agakishiev:2011pe} ($0.15 < p_T < 4$~GeV, as in the experiment). 
\label{fig:v2n}} 
\end{figure*}  

To illustrate the relevance of the effect, in Fig.~\ref{fig:C02} we show the results of our simulations for 
several cases for the like-sign $(++,--)$ and unlike-sign $(+-)$ pairs. In panel (a) 
we show the correlation $C(++,--)$ without direct balancing. We note the completely flat ridges, reflecting the approximate 
boost-invariance in the 
investigated kinematic range and, of course, the presence of flow.  We use the framework of event-by-event viscous 
hydrodynamics which generates realistic elliptic and triangular flows in the collisions
 \cite{Schenke:2011zz,*Qiu:2011hf,*Alver:2010dn}.
Therefore the dominant modulation of the shape in azimuth of the elliptic and triangular flows is well reproduced 
\cite{Takahashi:2009na,Werner:2012xh}.
Panel~(b) shows the same for $C(+-)$, where some mild fall-off 
in $\Delta \eta$ of the same-side 
ridge follows from the resonance decays. Panels~(c) and (d) include the direct charge balancing. 
We now note a prominent fall-off of the same-side ridge in
$C(+-)$, which is our key observation: the quantity $C(+-)-1$ drops from the central region to $|\Delta \eta| =2$ by about 
a factor of 2. The fall-off is also enhanced for $C(++,--)$ 
due to secondary effects from balancing of heavier particles, which later decay. The results of panels (c,d) are in qualitative and 
approximate quantitative agreement to the results of the STAR Collaboration, where the HBT correlations 
for identical particles are subtracted~\cite{Agakishiev:2011pe}.

To check whether our mechanism is correct also at the quantitative level, we now proceed to the investigation 
of the {\em charge balance functions}, defined 
as \mbox{$B(\Delta \eta)=\langle \ N_{+-}-N_{++} \rangle / \langle N_+ \rangle + \langle \ N_{-+}-N_{--} \rangle / \langle N_- \rangle$}, 
where $\langle N_{ab} \rangle$ denotes the event-averaged distributions of particles $a$ and $b$ with relative rapidity $\Delta \eta$, and 
$\langle N_a \rangle$ stands for the average number of particles $a$ in the acceptance window $|\eta| < 1$. 
We note that the charge balance function is related to the distributions in the numerator of Eq.~(\ref{eq:def}), 
\begin{eqnarray}
B(\Delta \eta) & = & \frac{\int d\Delta\phi [N^{\rm pair}_{+-}(\Delta \eta, \Delta \phi)-N^{\rm pair}_{++}(\Delta \eta, \Delta \phi)]}{2\pi \langle N_+ \rangle } +
\nonumber \\ &&  ( + \leftrightarrow -). \label{eq:bal}
\end{eqnarray}
The outcome, with correct agreement to the data, is presented in Fig.~\ref{fig:bal}. 
We note a preference to the lower freeze-out temperature, $T_f=140$~MeV.

The next quantitative investigation concerns the dependence of the flow coefficients on $\Delta\eta$, defined as 
\begin{eqnarray}
v_n^2(\Delta \eta) = \int d\Delta\phi \, \cos(n\Delta\phi) C(\Delta \eta, \Delta \phi). \label{eq:vn2}
\end{eqnarray}
The projection on the $\Delta \eta$ axis of the different harmonics  yields the squares of the consecutive  flow components $v_n$ present in the dihadron
correlation functions. 
The results presented in Fig.~\ref{fig:v2n} show agreement with the experiment, best for the mid-peripheral collisions and $T_f=140$~MeV. 
For the peripheral collisions, where the hydrodynamic approach is less justified, the agreement is qualitative, indicating 
that the hydrodynamic calculation overestimates the elliptic flow for large centralities. 
The experimental bands are extracted by integrating a model function fit to the measured dihadron correlations 
\cite{Agakishiev:2011pe}, varying the fit parameters within the estimated uncertainty. 
We note that these uncertainties are large for the central and peripheral cases. 
Our simulations incorporating the direct charge balancing (thick lines) exhibit the quested fall-off with $|\Delta \eta|$, while the cases without 
direct balancing (thin lines) are 
flat. The  independence of $v_n^2$ on $\Delta  \eta$ for the emission without charge balancing reflects the approximate pseudorapidity 
independence of the collective flow in the considered kinematic window. Charge balancing induces an additional component in $C$, of 
limited range $|\Delta \eta|\simeq 1$. The collimation of the opposite charge pairs occurs in the relative angle as well 
\cite{Bozek:2004dt,Aggarwal:2010ya}. As a result,
the contribution from charge balancing in $C(\Delta \eta,\Delta \phi)$ acquires the form of a 
2-dimensional peak at $\Delta \eta=\Delta \phi=0$.
The shape in $\Delta \eta$ of the  non-flow component  in $v_3^2$ is qualitatively reproduced in the simulations, but the 
overall strength is somewhat larger than extracted from the model fit in \cite{Agakishiev:2011pe}.
Thus our study shows that the charge balancing is the non-flow source of the observed $\Delta\eta$ dependence of the flow coefficients \cite{Aamodt:2011vk}.
The qualitatively similar behavior of higher-order harmonics, which needs higher statistics in our simulation,
as well as $v_1^2$, where the effects of the transverse-momentum conservation (not included in the present study) are 
important~\cite{Borghini:2000cm,*Bzdak:2010fd}, will be presented elsewhere. 
\begin{figure*}[tb]
\begin{center}
\includegraphics[width=0.35 \textwidth]{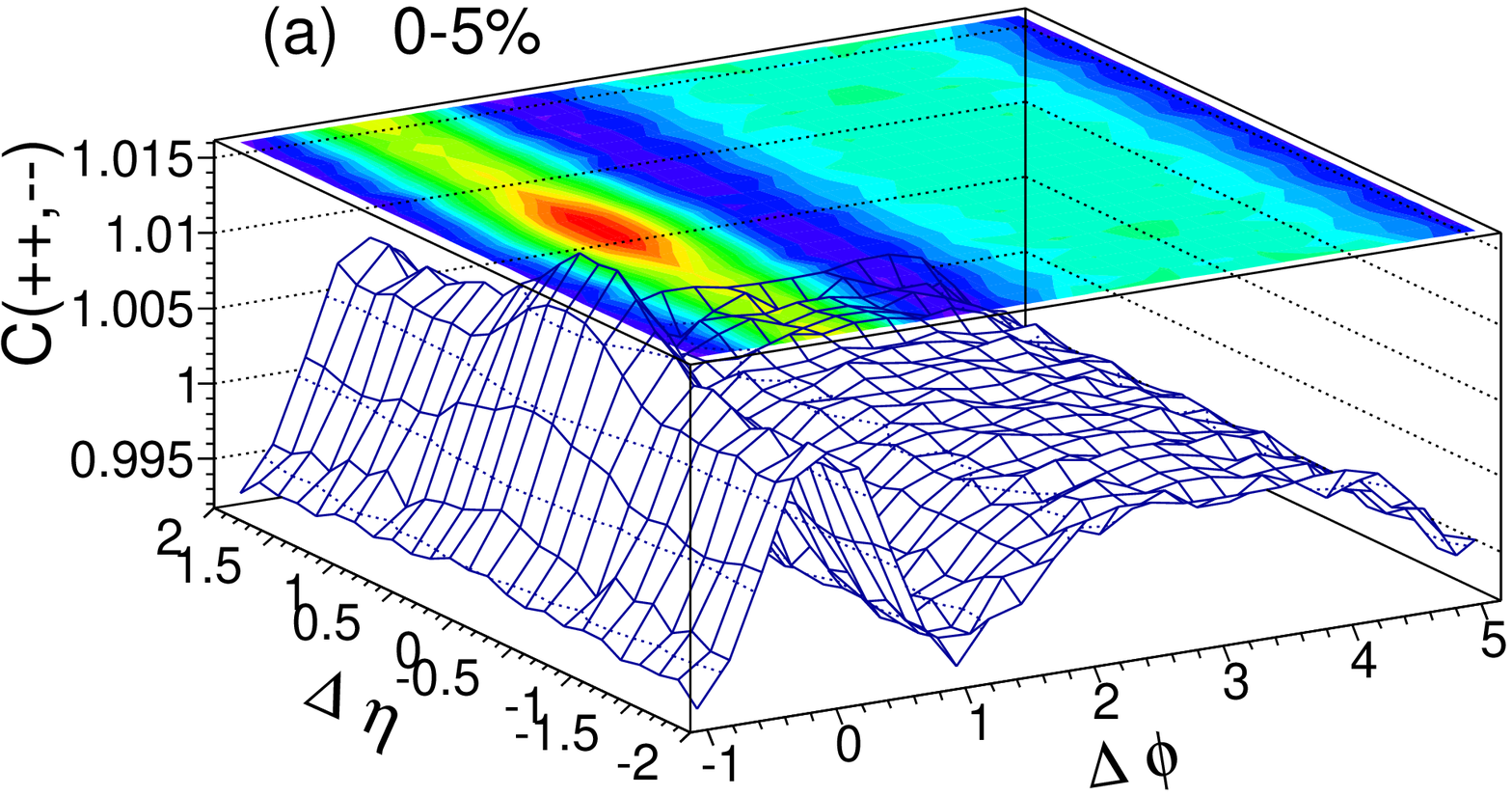} \hspace{-7mm}
\includegraphics[width=0.35 \textwidth]{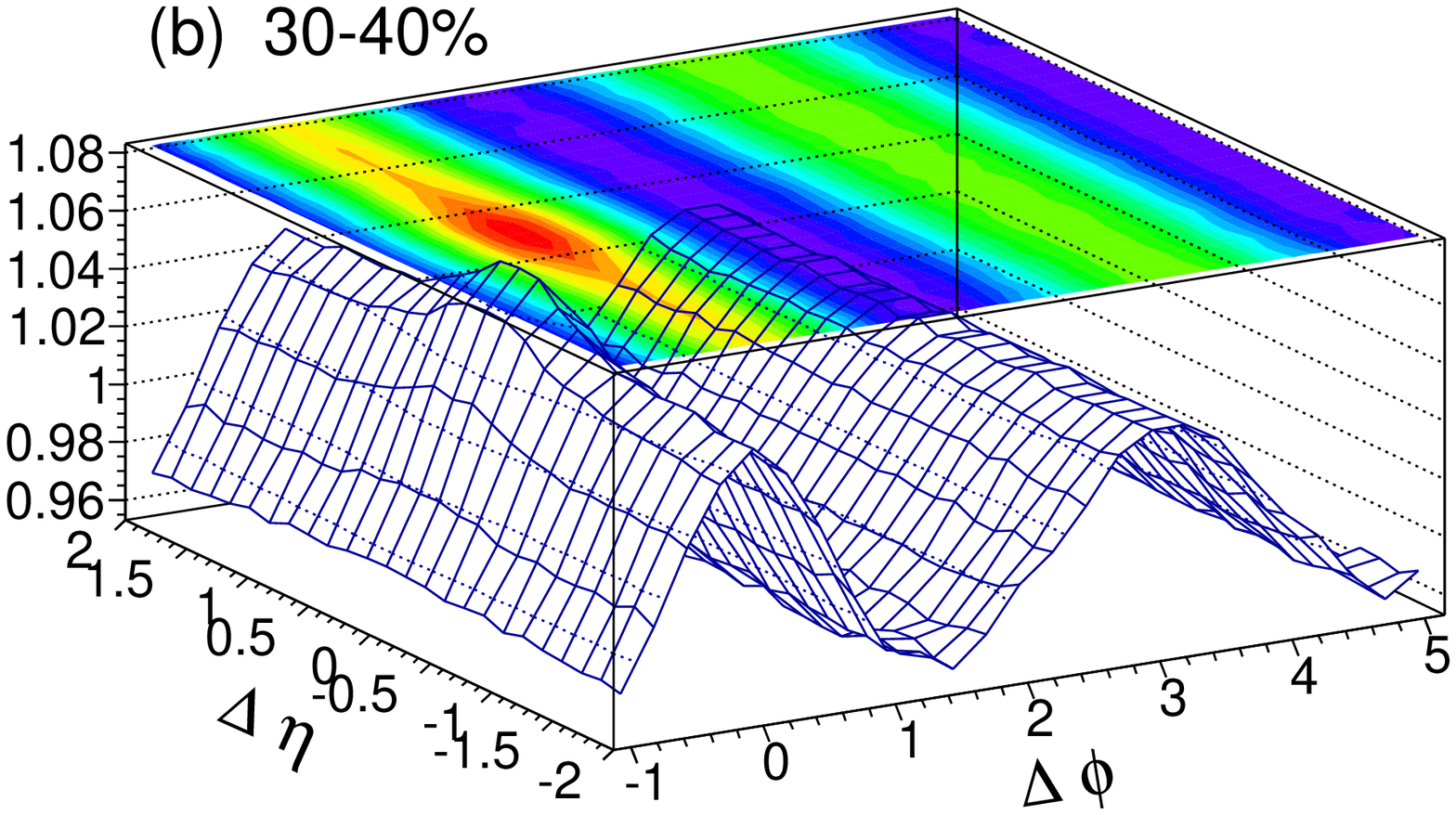} \hspace{-7mm}
\includegraphics[width=0.35 \textwidth]{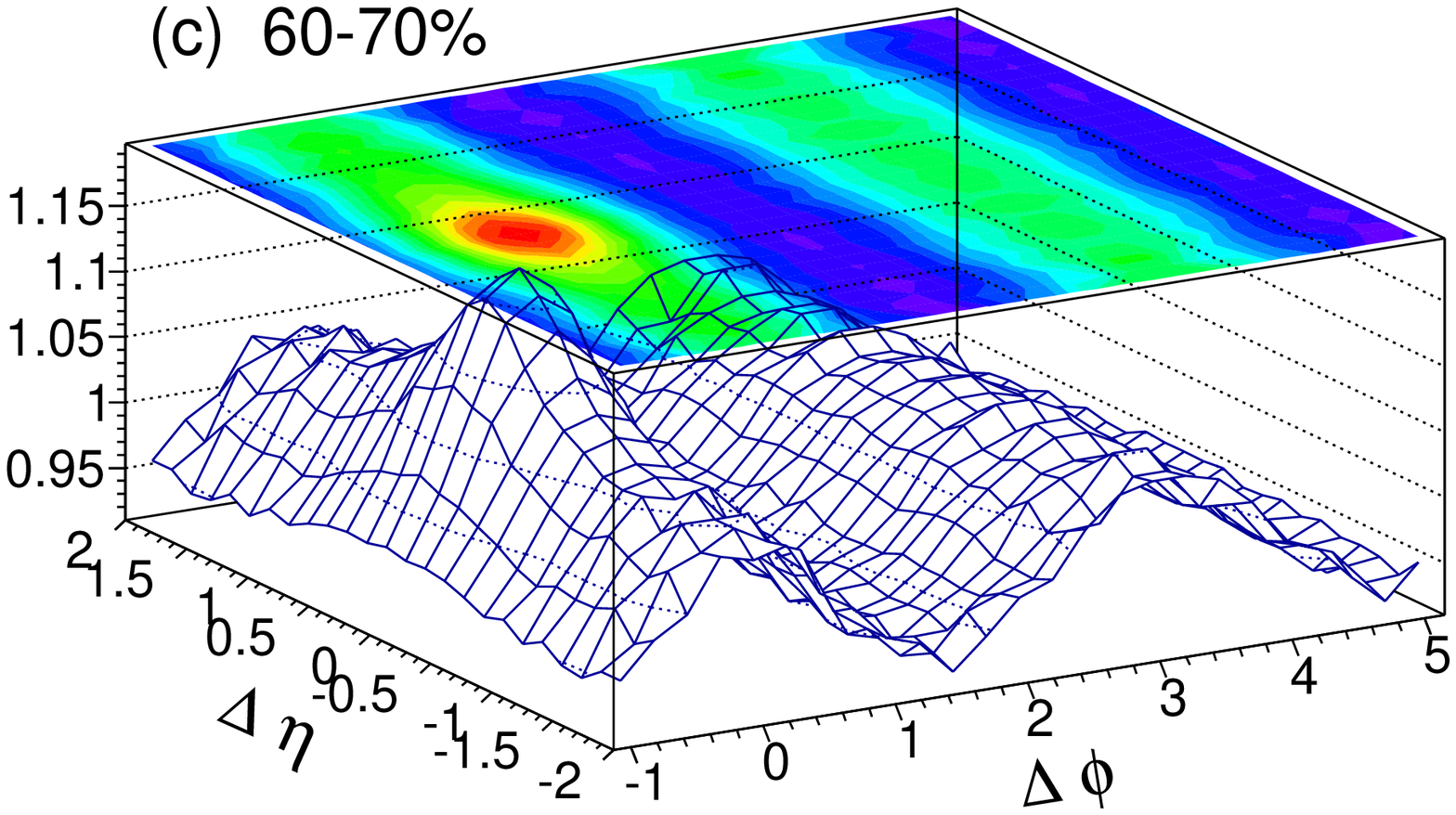} \\
\vspace{-7mm}
\includegraphics[width=0.35 \textwidth]{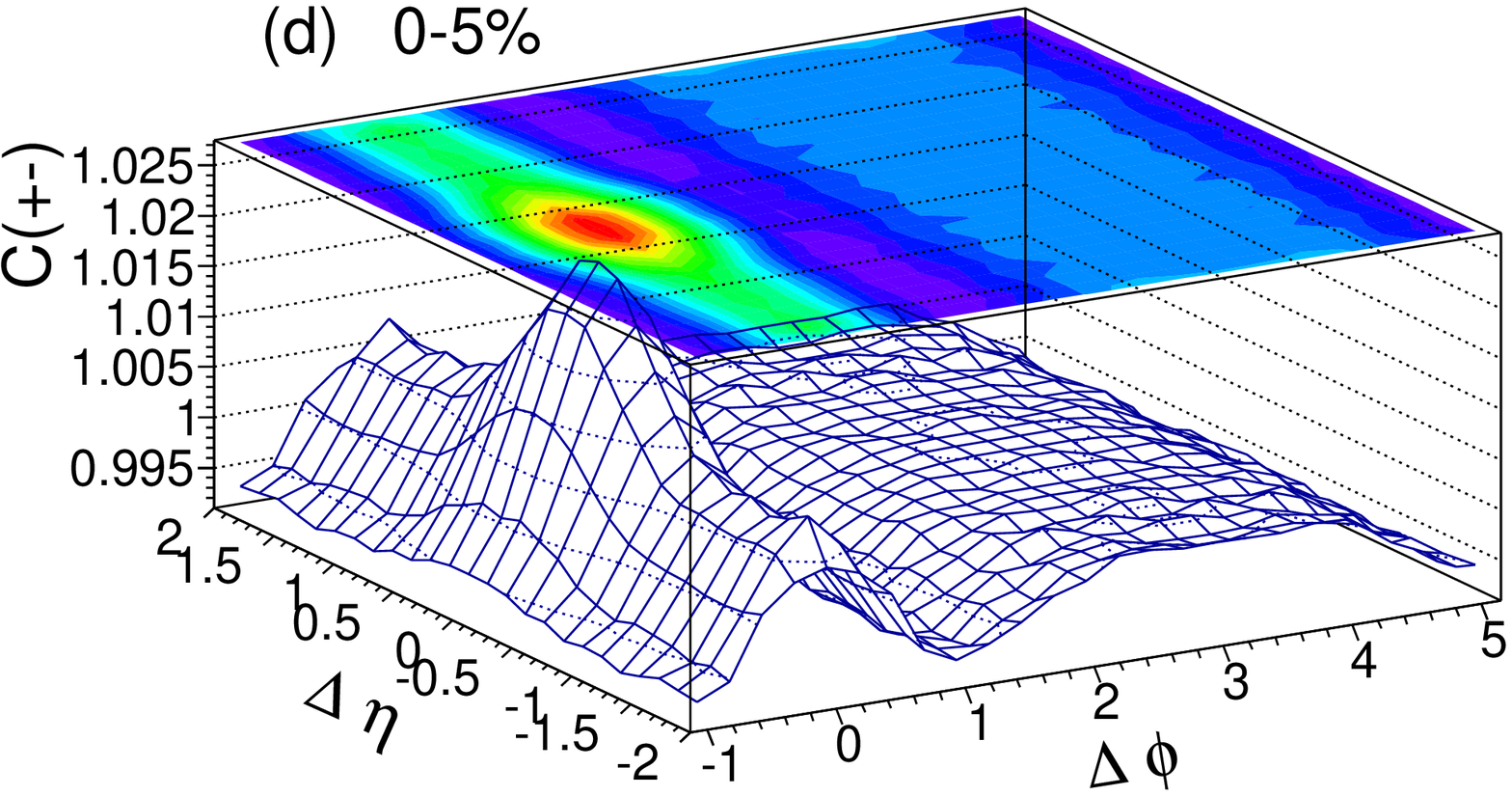} \hspace{-7mm}
\includegraphics[width=0.35 \textwidth]{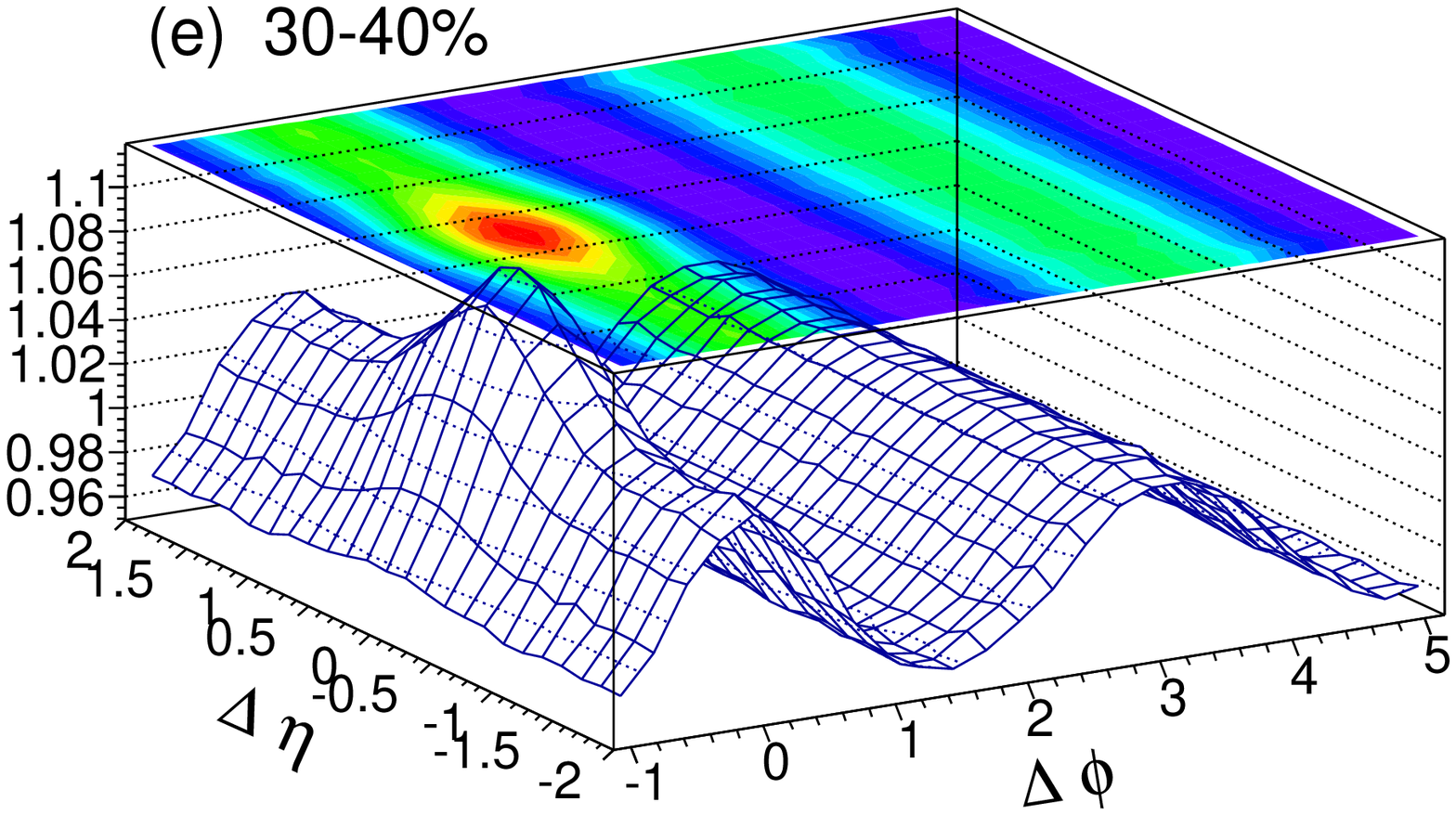} \hspace{-7mm}
\includegraphics[width=0.35 \textwidth]{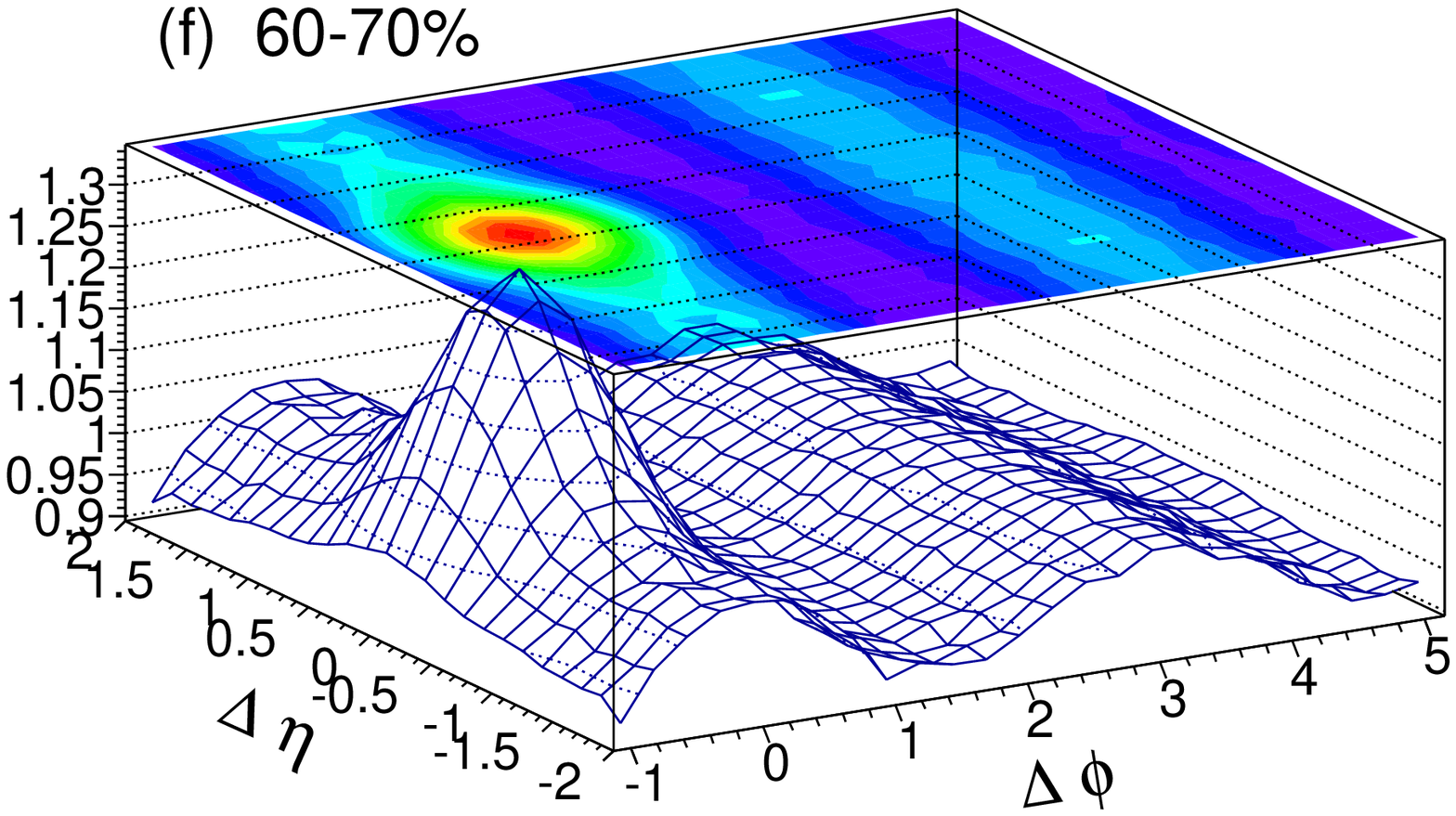} 
\end{center}
\vspace{-7mm}
\caption{(Color online) 
Our simulations for the correlation function $C$ with direct charge balancing 
included for the like-sign (a,b,c) and unlike-sign (d,e,f) pairs 
at three sample centralities ($T_f=140$~MeV, $0.8 < p_T < 4$~GeV as in Ref.~\cite{Abelev:2008un}).
\label{fig:C08}} 
\end{figure*}  

One may also compare the correlation function $C$ directly to the data shown, e.g., in Figs.~1 and~2 of Ref.~\cite{Abelev:2008un}, obtained for  
$0.8 < p_T < 4$~GeV, and with the HBT peak for the same-sign pairs removed. Our 
simulations shown in Fig.~\ref{fig:C08} display, for the first time in an approach based on hydrodynamics, all qualitative
features of the data and remain also in fair quantitative agreement. In particular, we note the proper dependence 
on the relative charge and centrality. Notably, the combinations $C(+-)-C(++,--)$ obtainable from Fig.~\ref{fig:C08} exhibit no ridges whatsoever, 
as they cancel out, leaving the central peak as the only structure. 

In conclusion, we remark that the presented simple effect based on the local charge conservation in the 
hadronization process is generic in its nature. It should manifest itself in all 
approaches where the charge balancing is combined with a collective motion of the source. 
Our approach, based on the fluctuating Glauber-model initial conditions, state-of-the-art hydrodynamics, and
statistical hadronization incorporating the direct charge balancing, is capable of reproducing all basic features of the 
data for the unbiased correlation function $C(\Delta\eta, \Delta\phi)$, as well as for the related quantities, such as the charge balance 
function and the harmonic flow coefficients $v_n^2(\Delta\eta)$. The correlation from charge balancing, 
yielding a two-dimensional central peak, comes on top of the ridge structures following from the presence of the 
azimuthally asymmetric collective flow~\cite{Takahashi:2009na,Luzum:2010sp}.
It thereby brings in a crucial non-flow component in the harmonic flow coefficients $v_n^2$, with a characteristic fall-off in 
the relative pseudorapidity. Thus the {\em collective flow} together with the {\em local charge conservation} 
is the key to a successful explanation of the shape of the correlation data in relativistic heavy-ion collisions. 


We thank Paul Sorensen for a discussion concerning the fall-off of the ridge, and to Miko\l{}aj Chojnacki 
for help in modifying {\tt THERMINATOR}. Supported by Polish Ministry of Science and Higher Education, grant N~N202~263438, and National Science 
Centre, grant DEC-2011/01/D/ST2/00772. The simulations were partly carried out on the Cracow Cloud One cluster.


%

\end{document}